\documentclass{kluwer} 
\usepackage{epsfig,psfig}
\def\ga{\lower.4ex\hbox{$\;\buildrel >\over{\scriptstyle\sim}\;$}}
\def\la{\lower.4ex\hbox{$\;\buildrel <\over{\scriptstyle\sim}\;$}}

\def\be{\begin{equation}}
\def\ee{\end{equation}}
\def\bea{\begin{eqnarray}}
\def\eea{\end{eqnarray}}

\def\half{{\textstyle{1\over2}}}

\begin{document}                                                                                   

\begin{opening}     
\title{Conservation of both current and helicity\\ in a quadrupolar model for solar flares}

\author{Don \surname{Melrose}}
\runningauthor{Don Melrose}
\runningtitle{Current and helicity conservation}
\institute{School of Physics,
University of Sydney, NSW 2006, AUSTRALIA}
\date{\today}

\begin{abstract}
A model for a solar flare, involving magnetic reconnection transferring flux and current between current-carrying magnetic loops connecting two pairs of footpoints, is generalized to include conservation of magnetic helicity during reconnection, as well as conservation of current at all four footpoints. For a set of force-free loops, with the $i$th loop having flux $F_i$ and current $I_i$, the self and mutual helicities are proportional to the self and mutual inductances with the constant of proportionality determined by $\alpha_i=F_i/\mu_0I_i$. In a constant-$\alpha$ model, the change in magnetic energy is proportional to the change in helicity, and conservation of helicity implies conservation of magnetic energy, so that a flare cannot occur. In a quadrupolar model,  with $\alpha_1>\alpha_2$ initially, $\alpha_1$ increases and $\alpha_2$ decreases when flux and current are transferred from loops~1 and~2 to loops~3 and~4. A model that conserves both current and helicity is constructed; it depends on the initial $\alpha$s, and otherwise is somewhat simpler than when helicity is neglected.
\end{abstract}

\end{opening}         

\section{Introduction}
Solar flares are attributed to magnetic energy release in the solar corona. A generic flare model involves a pre-flare magnetic configuration relaxing through magnetic reconnection to a post-flare configuration with a lower energy, with the magnetic energy difference identified as the energy released in the flare. However, the data on pre- and post-flare magnetic configurations are far from definitive in determining the details of the change in the magnetic configuration. As a consequence all detailed flare models rely on additional assumptions to constrain the pre- and post-flare configurations. One class of model involves reconnection between two current-carrying magnetic flux loops \cite{Metal88,Ketal95,Wetal95,Uetal03}. Such models are referred to as `two-loop' models, although four loops connecting the two pairs of footpoints are actually involved; the alternative name `quadrupolar' \cite{Uetal03} is used here. In such a model, magnetic reconnection involves transfer of magnetic flux and current between two pairs of loops. During a flare, the magnetic configuration in the corona changes, but there is insufficient time for the magnetic field below the photosphere to change. This is due to the change in magnetic stress propagating at the Alfv\'en speed, with the Alfv\'en propagation time being shorter than the flare time in the corona but longer than the flare time for propagation to any significant depth below the photosphere. This leads to the important constraint that the flux and current passing through the photosphere at each footpoint is unchanged during a flare. This constraint rules out several otherwise possible models, including a global change from higher to lower $\alpha$ and annihilation of two oppositely directed currents.

The non-potential component of the magnetic energy, due to the currents flowing in the corona, is described in terms of the currents and the self and mutual inductances. Current conservation implies that the changes that lead to magnetic energy release must be in the configuration of the current, rather than changes in the magnitude of the current. In a force-free corona, the current paths are along magnetic field lines. A simplifying assumption that allows one to model the change in the current configuration is that there are fixed current paths (loops) in the corona, with a fixed set of self and mutual inductances describing these paths. The changes during a flare correspond to current being transferred between these paths \cite{M97}. The model involves a large number of parameters: ten (four self inductances, six mutual inductances) describing the geometry, four describing the initial currents in the four loops, and two describing the current and flux transferred. Simplifying assumptions are made to reduce the number of free parameters, allowing one to identify the most favorable configurations for a flare with the maximum magnetic energy difference between the initial and final configurations \cite{HMH98}. Favorable configurations for magnetic energy release implied by this model are found to be correspond to favorable configurations found observationally for flares \cite{Aetal99}.

Magnetic helicity is conserved during reconnection \cite{T86}, and such conservation is believed to apply to solar flares \cite{PCZ96,B99,P99}. Conservation of helicity implies that helicity transported into the corona can only be rearranged in the corona, and must ultimately be transported out of the corona. Within the corona helicity can only be transferred from one set of field lines to another, from one scale to another or between twist and writhe \cite{YB03}. The transport of helicity into and out of the corona has received considerable attention recently, both for injection through the photosphere \cite{BR00, Ketal02,DB03,ML03,WL03,ZL03} and ejection through CMEs(Rust 1994, 1999; Kuijpers 1997) or back through the photosphere \cite{vBM89}. The time for transport of helicity out of the corona is long compared with the reconnection timescale, and hence helicity should be conserved on the flare timescale. Consequently, release of energy during a flare requires that the magnetic configuration change from one of higher magnetic energy to one of lower magnetic energy, with the net helicity of both configurations being the same. 

In this paper conservation of magnetic helicity is included in the reconnecting loop model. This provides an additional constraint on the flare model: magnetic flux and current at the footpoints, and helicity in the corona are all assumed to be conserved between the pre- and post-flare states. Note that conservation of helicity in a flare does not necessarily imply that the helicity is conserved between the observed pre- and post-flare configurations in the corona. For example, in eruptive flares the helicity carried off by CMEs \cite{R99} needs to be included in the helicity budget. 

The requirement of helicity conservation is included in a force-free multi-circuit model in section~2. The quadrupolar flare model is summarized in section~3. In section~4, implications of conservation of helicity are considered in a model for differential flux and current transfer are considered. The model is generalized to an arbitrary transfer of flux and energy in section~5. The results are summarized and discussed in section~6.

\section{Multiple circuit approximation}

The magnetic energy may be written in terms of the current. One starts from the general form
\be
E_{\rm mag}={1\over2}\int d^3{\bf x}\,{\bf J}({\bf x})\cdot{\bf A}({\bf x}),
\label{Edef}
\ee
where ${\bf J}({\bf x})$ is the current density, and writes the vector potential (in the Coulomb gauge) in terms of the current by solving Poisson's equation:
\be
{\bf A}({\bf x})=\mu_0\int d^3{\bf x}'\,
{{\bf J}({\bf x}')\over|{\bf x}-{\bf x}'|}.
\label{Ax}
\ee
The resulting double integral,
\be
E_{\rm mag}={\mu_0\over2}\int d^3{\bf x}d^3{\bf x}'\,
{{\bf J}({\bf x})\cdot{\bf J}({\bf x}')\over|{\bf x}-{\bf x}'|}
\label{Emag1}
\ee 
is the basis for a multiple circuit model (e.g., Jackson 1975). The current configuration is approximated by a set of discrete currents, $I_i$ with $i=1,2,\ldots$, and the integral (\ref{Emag1}) is approximated by
\be
E_{\rm mag}=\half\sum_{ij}M_{ij}I_iI_j=\half\sum_iL_iI_i^2+\sum_{i<j}M_{ij}I_iI_j,
\label{Ecm}
\ee
where $L_i=M_{ii}$ is the self-inductance for the $i$th current and $M_{ij}=M_{ji}$ for $i\ne j$ is the mutual inductance between the $i$th and $j$th currents. By choosing a sufficiently large number of discrete currents, the multiple circuit model may be used to obtain approximations of arbitrary accuracy. However, the model is useful in the present context only if the configuration of interest can be approximated by a small number of discrete currents. 

The magnetic helicity, $H$, is defined by
\be
H=\int d^3{\bf x}\,{\bf B}({\bf x})\cdot{\bf A}({\bf x}),
\label{Hdef}
\ee
and it may be expressed in terms of the current using (\ref{Ax}) twice, with ${\bf B}({\bf x})={\rm curl}\,{\bf A}({\bf x})$. A simplifying assumption made here is that the only contribution of interest is for force-free configurations of the magnetic field. This corresponds to
\be
{\bf J}({\bf x})={\alpha({\bf x})\over\mu_0}\,{\bf B}({\bf x}),
\label{ffc}
\ee
with $\alpha({\bf x})$ constant along each current line, which coincides with the magnetic field line. Then the integral (\ref{Hdef}) with (\ref{ffc}) is of the same form as the energy integral (\ref{Edef}). In the multiple circuit model the relation (\ref{ffc}) implies a relation
\be
I_i={\alpha_i\over\mu_0}\,F_i
\label{ffc2} 
\ee
between the current and the magnetic flux associated with the $i$th loop, and where $\alpha_i$ is an appropriate average of $\alpha({\bf x})$ over the cross-section of the loop. Then the analogy with the magnetic energy implies that the helicity may be written in the form
\be
H=\sum_{ij}m_{ij}I_iI_j
=\sum_il_iI_i^2+\sum_{i<j}2m_{ij}I_iI_j,
\label{Hcm1}
\ee
with the term involving $l_i=m_{ii}$ describing the self helicity and the terms involving $m_{ij}=m_{ji}$ describing the mutual helicity. These terms are related to the self inductances $L_i$ and the mutual inductances $M_{ij}$ by
\be
l_i={\mu_0L_i\over\alpha_i},
\qquad
m_{ij}={\mu_0M_{ij}\over\alpha_{ij}},
\qquad
\alpha_{ij}={2\alpha_i\alpha_j\over\alpha_i+\alpha_j},
\label{Hij1}
\ee
where the symmetry property $m_{ij}=m_{ji}$ is imposed without loss of generality.

Alternatively, the expression (\ref{Hcm1}) for $H$ may be expressed in terms of the fluxes using (\ref{ffc2}). This gives
\be
H=\sum_{ij}{\cal L}_{ij}F_iF_j
=\sum_iT_iF_i^2+\sum_{i<j}2{\cal L}_{ij}F_iF_j,
\label{Hcm2}
\ee
with the term $T_i={\cal L}_{ii}$ describing the self helicity and the terms involving ${\cal L}_{ij}={\cal L}_{ji}$ describing the mutual helicity. The form (\ref{Hcm2}) for $H$ was written down by Berger \shortcite{B98,B99}. The coefficients in (\ref{Hcm2}) are related to the self inductances $L_i$ and the mutual inductances $M_{ij}$ by
\be
T_i={\alpha_i\over\mu_0}\,L_i,
\qquad
{\cal L}_{ij}={\alpha_i+\alpha_j\over\mu_0}\,M_{ij}.
\label{Hij2}
\ee
The quantity $T_i$ may be interpreted as the number of twists (angle of the twist divided by $2\pi$) along the $i$th loop.

Although explicit expressions for the coefficients $T_i$ and ${\cal L}_{ij}$ have been presented in the literature, these are not satisfactory for present purposes. There are several properties that an acceptable form of ${\cal L}_{ij}$ should satisfy: (i) ${\cal L}_{ij}={\cal L}_{ji}$ should be symmetric in interchange of the loops (any asymmetric part has no physical significance), (ii) ${\cal L}_{ij}$ should have its maximum  positive value when the loops are identical, parallel and coincident, when ${\cal L}_{ij}$ should be equal to $T_i=T_j$ (the currents and fluxes of the two loops then add, (iii) ${\cal L}_{ij}$ should have its maximum negative value when the loops are identical, antiparallel and coincident, when ${\cal L}_{ij}$ should be equal to $-T_i=-T_j$ (the combination of the two loops is null), and (iv) ${\cal L}_{ij}=0$ when the two loops are orthogonal. The available calculations of the mutual helicity involve first showing that the rate of change of the mutual helicity between loops (flux tubes) $i$ and $j$ is
\be
{dH_{ij}\over dt}=-{2\over\pi}\,\Omega_{ij}\,F_iF_j,
\label{dHijdt}
\ee
where $\Omega_{ij}$ is `the angular velocity of the $j$th flux tube about the $i$th' \cite{WL03}, and then integrating from some specific motion starting from an initial condition where $H_{ij}$ is zero \cite{B98}. The resulting expression \cite{B98} appears to satisfy none of the requirements (i)--(iv). This difficulty is implicit in the lack of symmetry between $i$, $j$ in the interpretation of $\Omega_{ij}$ in (\ref{dHijdt}), and in the lack of symmetry in the manner in which the integral over time is performed. In a detailed theory, in which an explicit expression for the mutual helicity is required, this difficulty must be addressed explicitly. However, the arguments in the present paper do not rely on any explicit expression for ${\cal L}_{ij}$, and this difficulty is not discussed further here as it does not affect any of the conclusions.

Nevertheless it is appropriate to identify an explicit expression for the mutual helicity that satisfies the requirements (i)--(iv). A semi-empirical result follows by modifying an approximate expression for the mutual inductance \cite{M97}, which depends only on the geometry of the current paths. This semi-empirical result is
\be
{\cal L}_{ij}=(T_iT_j)^{1/2}\,\cos\theta_{ij}\,
\left({4a_ia_j\over(a_i+a_j)^2+d_{ij}^2}\right)^{3/2}.
\label{Lij}
\ee
where $\theta_{ij}$ is the angle between loops~$i$ and~$j$, such that $\theta_{ij}=0$ when the loops are parallel, $\theta_{ij}=\pi$ when they are antiparallel, and  $\theta_{ij}=\pi/2$ when they are orthogonal, and where $a_i$, $a_j$ are the lengths of the loops (the major radii in a toroidal model) and $d_{ij}$ is the distance between their centers. It is desirable that an expression for ${\cal L}_{ij}$ that includes the important features incorporated into (\ref{Lij}) be derived rigorously.

\begin{figure}
\label{figm97a}
\centerline{
\psfig{file=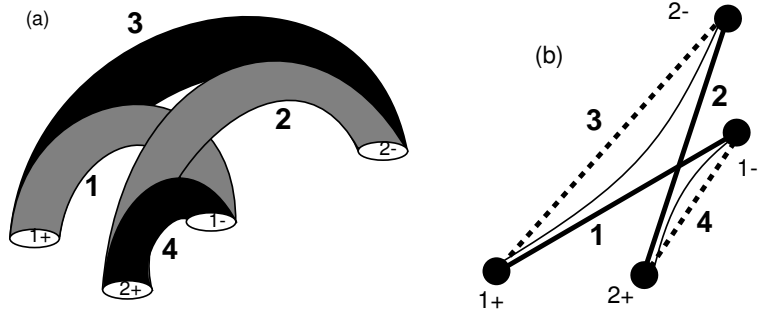,width=10cm}}
\caption{The quadrupolar model: (a) the two initial loops are shaded lightly and the two final loops are dark; (b) in projection on the photosphere, the initial loops are the solid lines, the final loops are the dotted lines, and the faint curves denote transferring flux and current. [After Melrose (1997)]}
\end{figure}

\section{A quadrupolar flare model without helicity  conservation}

The flare model adopted here \cite{M97}, cf.\ figure~1, is based on the change in the magnetic energy in the form (\ref{Ecm}) as the result of a change in the magnetic configuration during a flare. An important assumption in the model is that only the initial (pre-flare) and final (post-flare) states are of direct interest, and in both cases the energy may be written in terms of the currents and inductances. (The inductances are changing during the reconnection process, and an energy integral then does not exist in general.) 

The energy released is identified as
\be
E_{\rm fl}=E_{\rm mag}-E'_{\rm mag},
\qquad
E'_{\rm mag}=\half\sum_{ij}M'_{ij}I'_iI'_j,
\label{Efl}
\ee
where the unprimed and primed quantities denote the pre- and post-flare configurations, respectively. The assumption that the geometry does not change corresponds to $M'_{ij}=M_{ij}$. A flare is possible only if there is free energy to drive it, and this requires $E_{\rm fl}>0$. In principle, (\ref{Efl}) could be used to calculate the energy release as a result of any change in the magnetic structure and current distribution in the corona, simply by modeling the pre- and post-flare fields with as many loops as desired. 

A generic model involves transfer of flux and current between pairs of footpoints. Let the $i$th pair of footpoints be denoted $i_\pm$. A particular loop connects two footpoints, and let the loops  be denoted by the footpoints that they connect. Thus loop $i_+j_-$ connects footpoints $i_+$ and $j_-$, and has a magnetic flux $F_{i_+j_-}$ and a current $I_{i_+j_-}$. Both flux and current must be conserved at each footpoint in such a model. Conservation of current at the footpoint $i+$, say, requires
\be
I_{i_+}=I'_{i_+},
\qquad
I_{i_+}=\sum_{j_-}I_{i_+j_-},
\qquad
I'_{i_+}=\sum_{j_-}I'_{i_+j_-},
\label{Ii+}
\ee
and conservation of magnetic flux requires
\be
F_{i_+}=F'_{i_+},
\qquad
F_{i_+}=\sum_{j_-}F_{i_+j_-},
\qquad
F'_{i_+}=\sum_{j_-}F'_{i_+j_-},
\label{Phii+}
\ee
where the sums are over all the footpoints $j_-$ connected to $i_+$ by a loop. There is one such constraint for each footpoint.

In practice, we have no direct information on the reconfiguration of the magnetic field, and we need to make major simplifying assumptions to formulate a flare model. The assumptions made here are:
\begin{description}
\item[(a)] The current paths (loops) are predetermined, and the changes involve redistributing the current between these paths. In other words, it is assumed that the transfer of flux and current is between loops with a fixed geometry and this geometry (which determines the inductances) does not change as a consequence of this transfer, so that one has $M'_{ij}=M_{ij}$.

\item[(b)] A flare is modeled in terms of current and flux transfer between two pairs of footpoints, so that only four loops are involved.
\end{description}
Note that in the quadrupolar model there are four loops, thought of as two initial loops and two additional final loops formed by making new magnetic connections between the four footpoints. Let loops $1_+1_-$. $2_+2_-$, $1_+2_-$, $2_+1_-$ be labeled more simply as loops 1, 2, 3, 4, and let the initial and final currents flowing in these loops be denoted by unprimed and primed quantities, respectively. It is assumed that a current $\Delta I$ and a flux $\Delta F$ are transferred from the initial loops (1 and 2) to the final loops (3 and 4). Conservation of current implies $I'_1=I_1-\Delta I$, $I'_2=I_2-\Delta I$, $I'_3=I_3+\Delta I$, $I'_4=I_4+\Delta I$, and conservation of flux implies an analogous set of relations. It is convenient to write the energy released in the flare as \cite{M97}
\be
E_{\rm fl}=R\Delta I + M^{\rm IR}(\Delta I)^2,
\label{m97a}
\ee
with the coefficients given by
\bea
R=M_1^{\rm LCS}I'_1+M_2^{\rm LCS}I'_2
-M_3^{\rm LCS}I_3-M_4^{\rm LCS}I_4,
\nonumber
\\
\noalign{\vskip3pt}
M_j^{\rm LCS}=M_{j1}+M_{j2}-M_{j3}-M_{j4},
\qquad\qquad
\\
\noalign{\vskip3pt}
M^{\rm IR}=\half(L_1+L_2-L_3-L_4)+M_{12}-M_{34},
\qquad
\label{m97b}
\eea
where LCS labels `like current separation' terms and IR labels `irreducible reconnection' terms. 

\section{Infinitesimal change conserving current and helicity}

It is helpful to consider the case of an inifinitesimal reconnection ($\Delta I\to dI$, $\Delta F\to dF$) that conserves both current and helicity. This implies a change in the helicity due to a change in the currents, $I_i$, rather than to a change in the geometry of the current (or magnetic field) pattern, as assumed in (\ref{dHijdt}). 

\subsection{Change in the magnetic energy}

An infinitesimal change $I_i\to I_i+dI_i$ in the currents implies an infinitesimal change in the magnetic energy:
\be
dE_{\rm mag}=\sum_{ij}M_{ij}I_idI_j.
\label{dEmag}
\ee
Conservation of current in the quadrupolar model requires $dI_1=dI_2=-dI_3=-dI_4=-dI$. The flux also changes, with $dF_1=dF_2=-dF_3=-dF_4=-dF$. The (\ref{dEmag}) becomes
\be
dE_{\rm mag}=-\sum_{i}(M_{i1}I_1+M_{i2}I_2-M_{i3}I_3-M_{i}I_4)dI.
\label{dEmag1}
\ee
One requires $dE_{\rm mag}<0$ for magnetic energy to be released to drive a flare.

\subsection{Change in the $\alpha$s}

Suppose one writes
\be
{dI\over dF}={\alpha_0\over\mu_0}.
\label{alpha0}
\ee
Equal amounts of current and flux are transferred from the two initial loops to the final loops, and if the initial loops have different $\alpha$s, then one expects the value of $\alpha_0$ to be a weighted average of $\alpha_1$ and $\alpha_2$. Moreover, if the two final loops are created by the reconnection then one expects them to have $\alpha_3=\alpha_4=\alpha_0$; this is because the same ratio ($\alpha_0/\mu_0$) of current to flux is transferred from both initial loops to both final loops.

Assuming $\alpha_1>\alpha_0>\alpha_2$, a notable consequence is that $\alpha_1$, $\alpha_2$ do not remain constant. Writing $\mu_0(I_i+dI_i)=(\alpha_i+d\alpha_i)(F_i+dF_i)$, one finds
\be
d\alpha_i=\alpha_i\left(1-{\alpha_i\over\alpha_0}\right){dI_i\over I_i}.
\label{dalphai}
\ee
In particular, one has $d\alpha_1=-\alpha_1(1-\alpha_1/\alpha_0)dI/I_1>0$, and $d\alpha_2=-\alpha_2(1-\alpha_2/\alpha_0)dI/I_2<0$. The loop with the larger (smaller) $\alpha$ has its $\alpha$ increased (decreased).  Superficially it might appear that this result violates Taylor's theorem, which implies that the system should relax to one of constant $\alpha$. In fact, the overall mean value of $\alpha$ does decrease as a result of the current and flux transfer.

\subsection{Infinitesimal change in the helicity}

The change in the $\alpha$s implies that the coefficients $m_{ij}$ in the expression (\ref{Hcm1}) for the helicity change. Using (\ref{Hij1}), with $dM_{ij}=0$ by hypothesis, the change in the helicity becomes
\be
dH=\sum_{ij}\mu_0M_{ij}\left({1\over\alpha_0}+{1\over\alpha_i}\right)I_idI_j,
\label{dH}
\ee
where (\ref{dalphai}) is used. 

The change in the energy (\ref{dEmag}) is subject to the condition $dH=0$ with $dH$ given by (\ref{dH}). Consider the identity $dE_{\rm mag}=dE_{\rm mag}+CdH$, which is satisfied for all constants $C$. Choosing $C$ such that the condition $dE_{\rm mag}=0$  is manifestly satisfied for constant-$\alpha$, that is for $\alpha_i=\alpha_0$ for all $i$, corresponds to $C=-\alpha_0/2\mu_0$. Then in place of (\ref{dEmag}) one has
\be
dE_{\rm mag}=\sum_{ij}{\alpha_i-\alpha_0\over2\alpha_i}M_{ij}I_idI_j.
\label{dEmag2}
\ee
Assuming $\alpha_3=\alpha_4=\alpha_0$, in place of (\ref{dEmag1}) one has
\be
dE_{\rm mag}=-\sum_{i}\left({\alpha_1-\alpha_0\over2\alpha_1}M_{i1}I_1
+{\alpha_2-\alpha_0\over2\alpha_2}M_{i2}I_2\right)dI.
\label{dEmag3}
\ee
Note that one can rewrite (\ref{dEmag3}) in term of the fluxes: $I_1/\alpha_1=F_1/\mu_0$, $I_2/\alpha_2=F_2/\mu_0$, $dI/\alpha_0=dF/\mu_0$.  

\section{Flare model that conserves helicity}

The foregoing discussion of the differential case suggests how to proceed in the more general case where there is an arbitrary transfer of current, $\Delta I$, and flux, $\Delta F=\mu_0\Delta I/\alpha_0$.  

\subsection{Generalization of the quadrupolar model}

Provided thet helicity is conserved one may introduce an equivalent energy by adding an arbitrary constant, $C$, times the helicity to the energy: then the change in the equivalent energy is equal to the change in the energy. As in the differential case it is appropriate to choose the constant  $C=-\alpha_0/2\mu_0$ in defining the equivalent energy:
\be
{\tilde E}_{\rm mag}=E_{\rm mag}-{\alpha_0\over2\mu_0}H,
\qquad
E_{\rm fl}={\tilde E}_{\rm mag}-{\tilde E}'_{\rm mag},
\label{flare1}
\ee
where conservation of helicity implies that the additional term $\propto H$ does not affect the value of the energy released. The choice (\ref{flare1}) is made to ensure that the requirement $E_{\rm fl}=0$ in a constant-$\alpha$ model be manifestly satisfied.

The change in the energy at constant helicity and constant  current then reduces to
\bea
E_{\rm fl}&=&{\Delta I\over4}\bigg\{
\left(1-{\alpha_0\over\alpha_1}\right)\big[
2L_1+M_{12}-M_{13}-M_{14}
\big]I_1
\nonumber\\
&&\qquad
+\left(1-{\alpha_0\over\alpha_2}\right)\big[
2L_2+M_{12}-M_{23}-M_{24}
\big]I_2
\nonumber\\
&&\qquad
-\left(1-{\alpha_0\over\alpha_3}\right)\big[
2L_3+M_{34}-M_{13}-M_{23}
\big]I_3
\nonumber\\
&&\qquad
-\left(1-{\alpha_0\over\alpha_4}\right)\big[
2L_4+M_{34}-M_{14}-M_{24}
\big]I_4
\bigg\}.
\label{flare2}
\eea

One may compare (\ref{flare2}), which incorporates helicity conservation, and (\ref{m97a}) with (\ref{m97b}), which does not include helicity conservation. For the purpose of the comparison one needs to make the replacements $I'_1\to I_1-\Delta I$,  $I'_2\to I_2-\Delta I$  in (\ref{m97a}) with (\ref{m97b}). Inspection then shows that (\ref{flare2}) is somewhat the simpler of the two expressions, in particular because there in no term $\propto(\Delta I)^2$. The way the helicity constraint is used, to require that the energy change be zero in the constant-$\alpha$ case, has the associated consequence of removing terms $\propto(\Delta I)^2$. In this sense, inclusion of the helicity constraint simplifies the model by relating the final configuration to the initial configuration, allowing one to eliminate explicit dependence on the final configuration.

\subsection{Model involving creation of final loops}

The simplification introduced by imposing conservation off helicity becomes more apparent when one considers the simpler case in which the two final loops are created by the reconnection. This model implies $I_3=I_4=0$,  $I'_3=I'_4=\Delta I$, and it also implies $\alpha_3=\alpha_4=\alpha_0$. Hence, the terms $\propto I_3,I_4$ drop out for two separate reasons ($I_3,I_4\to0$ and $\alpha_3=\alpha_4\to\alpha_0$). This suggests that the dependence of the result on the final configuration is now almost completely implicit in the helicity constraint.

It is further plausible to assume that the characteristic $\alpha_0$ of the flux transferred is the mean of the initial $\alpha$s:
\be
\alpha_0=\half(\alpha_1+\alpha_2).
\label{flare3}
\ee
With these assumptions (\ref{flare2}) simplifies to
\bea
E_{\rm fl}&=&{\Delta I(\alpha_1-\alpha_2)\over8}\bigg\{
\big[
2L_1+M_{12}-M_{13}-M_{14}
\big]{I_1\over\alpha_1}
\nonumber\\
&&\qquad\qquad\qquad
-\big[
2L_2+M_{12}-M_{23}-M_{24}
\big]{I_2\over\alpha_2}
\bigg\}.
\label{flare4}
\eea
The most favorable configuration for a flare is to be identified with the configurations that maximize $E_{\rm fl}$.

The sign of  $E_{\rm fl}$ is determined by the sign of $\alpha_1-\alpha_2$ and the sign of the quantity in curly brackets in (\ref{flare4}). For the terms involving the self inductances, which are the largest terms in many cases, to give $E_{\rm fl}>0$ requires $L_1I_1/\alpha_1>L_2I_2/\alpha_2$.  Hence, for $\alpha_1>\alpha_2$ the favorable requirement for a large energy release is $L_1I_1\gg L_2I_2$, that is, the dominant initial loop ($L_1I_1\gg L_2I_2$) needs to have the larger twist per unit length ($\alpha_1>\alpha_2$).

A simple model corresponds to all the current and flux in loop~2 being transferred to the final loops. This corresponds to $I_2=\Delta I$ and $\alpha_2=\alpha_0$. Neglecting the term $\propto I_2/\alpha_2$ in (\ref{flare4}), and maximizing the resulting expression at fixed $F_2=\Delta F$ implies $\alpha_2=\half\alpha_1$. Then (\ref{flare4}) simplifies to
\be
E_{\rm fl}={1\over16}\big[
2L_1+M_{12}-M_{13}-M_{14}
\big]\,I_1\,\Delta I.
\label{flare5}
\ee
Ignoring the mutual helicities, the change in the magnetic energy in loop 1 is $L_1I_1\Delta I$, and (\ref{flare5}) implies that one eighth of this energy becomes available as free energy to drive the flare. Comparison of this result with the corresponding result for this simple model for the case without the helicity constraint, one finds that (\ref{m97a}) with (\ref{m97b}) allows a greater fraction of the energy to be released as free energy. Hence, the helicity constraint precludes certain final magnetic configurations that might otherwise appear favorable for maximizing the energy release.

\section{Discussion and conclusions}

The primary objective of this paper is to include conservation of helicity in a quadrupolar flare model that is based on conservation of current and flux at two pairs of footpoints. To achieve this objective, the first step is to write the helicity in a form analogous to the energy, which is described by a collection of  loop currents and their self and mutual inductances. The corresponding self and mutual helicities are proportional to the inductances, with the constants of proportionality involving the $\alpha$s associated with the currents. 

Several implications of conservation of helicity are identified. First, in a constant $\alpha$ model, conservation of helicity implies conservation of magnetic energy, and hence a flare is impossible, subject to the proviso that the fluxes and currents at the four footpoints do not change. Second, conservation of helicity precludes a model in which transfer of current and flux from the two initial loops through reconnection creates two new final loops. Effectively, the final loops must already be present before current and flux can be transferred to them. How this difficulty is to be overcome in the model is not entirely clear. One suggestion is that the two new loops form by splitting the initial loops with the split parts moving apart into their final positions. Third, the expression of conservation of helicity may be used to rewrite the change in magnetic energy in a manner that explicitly satisfies the requirement that the change in energy be zero when all the $\alpha$s are the same. This results in an expression of the change in magnetic energy which is proportional to the difference in the initial $\alpha$s, but otherwise has a relatively simple form. In effect the constraint on the allowed final configurations imposed by helicity conservation allows one to write the change in a way that superficially does not depend on the final configuration. Fourth, a flare is allowed only if the stronger initial loop has the larger $\alpha$, and the $\alpha$ of the post-flare (reduced flux and current) loop is higher than the initial value.

The investigation in this paper is far from a complete discussion of the implications of conservation of helicity on the quadrupolar model. Three specific points clearly need further investigation. First, the application of the quadrupolar model to the interpretation of data on specific flares \cite{Aetal99} needs to be reconsidered, particularly in view of the importance of the ratio of the $\alpha$s in the two initial loops in the helicity-conserving form of the model. Second, a problem identified in the course of this investigation concerns explicit approximate expressions for the mutual helicity. Available expressions (Berger 1998, 1999; Priest 1999) do not satisfy four plausible requirements, (i)--(iv), discussed in section~2. A semi-empirical formula that satisfies these requirements is written down in (\ref{Lij}). Direct calculation of the mutual helicity needs to be reconsidered. Third, the model does not include magnetic structures that are unconnected to the photosphere, and such structures may form as a result of reconnection, and be identified with CMEs.

\acknowledgements
I thank Mike Wheatland for helpful discussions.

\end{document}